\documentclass[sigconf,nonacm]{acmart}
\usepackage{multirow}
\usepackage{tabularx} 
\usepackage{graphicx} 
\usepackage{makecell} 
\usepackage{booktabs}
\usepackage{amsmath}
\usepackage{array}
\AtBeginDocument{%
  }

\setcopyright{none}
\renewcommand\footnotetextcopyrightpermission[1]{}
\begin{document}

\title{Missing-Token Prompted Reliability-Aware Fusion for Robust Polyglot Speaker Identification}

\author{Peng Jia}
\authornote{Both authors contributed equally to this research.}
\email{2020214631@mail.hfut.edu.cn}
\affiliation{%
  \institution{Hefei University of Technology}
  \city{Hefei}
  \country{China}
}

\author{Li Dai}
\authornotemark[1]
\email{2025110483@mail.hfut.edu.cn}
\affiliation{%
  \institution{Hefei University of Technology}
  \city{Hefei}
  \country{China}
}


\author{Jia Li}
\authornote{Corresponding author.}
\email{jiali@hfut.edu.cn}
\affiliation{%
  \institution{Hefei University of Technology}
  \city{Hefei}
  \country{China}
}

\author{Zhenzhen Hu}
\email{zzhu@hfut.edu.cn}
\affiliation{%
    \institution{Intelligent Interconnected Systems Laboratory of Anhui Province}
  \institution{Hefei University of Technology}
  \city{Hefei}
  \country{China}
}


\author{Ye Zhao}
\email{zhaoye@hfut.edu.cn}
\affiliation{%
  \institution{Hefei University of Technology}
  \city{Hefei}
  \country{China}
}

\author{Richang Hong}
\email{hongrc.hfut@gmail.com}
\affiliation{%
  \institution{Hefei University of Technology}
  \city{Hefei}
  \country{China}
}



\begin{abstract}
Accurate and robust multimodal speaker identification is essential for multimedia understanding and biometric authentication. However, real-world polyglot scenarios pose two key challenges: speaker-discriminative representations should generalize across languages, and the model should remain reliable when face information is unavailable. To address these challenges, we propose \textbf{MRAF}, a \textbf{M}issing-Token Prompted \textbf{R}eliability-\textbf{A}ware \textbf{F}usion framework for polyglot speaker identification across complete-modality, missing-face, and cross-lingual scenarios. MRAF represents unavailable face inputs with a learnable missing token instead of fixed zero-valued features, providing a trainable representation of the missing visual state. This design reduces the distribution gap caused by missing inputs and allows subsequent reliability estimation and cross-modal fusion to operate within a unified token space. To adaptively integrate modalities with different reliability, MRAF further introduces a reliability-aware cross-attention fusion module, which estimates face and audio reliability scores, normalizes them into modality weights, and applies these weights to token representations before bidirectional cross-attention. In this way, the model can emphasize reliable modality cues while suppressing unreliable ones. During training, MRAF jointly optimizes multi-branch classification losses, audio-only knowledge distillation, and center loss to improve speaker discrimination and missing-modality robustness. Experiments on the official POLY-SIM 2026 test set demonstrate the effectiveness of the proposed framework. In the final evaluation, MRAF achieves 100\% accuracy on P3 and P5, obtains competitive results on the more challenging missing-face settings P4 and P6. The source code will be released at \url{https://github.com/MSA-LMC/MRAF}.
\end{abstract}




\maketitle

\section{Introduction}

Audio-visual speaker identification aims to recognize speaker identity by jointly modeling facial appearance and speech signals~\cite{roy2010introducing,nagrani2018seeing,horiguchi2018facevoice,nawaz2021cross}. Compared with unimodal systems, it exploits complementary cues from face and voice, making it useful for multimedia retrieval, biometric authentication, and human-centered video understanding. Large-scale in-the-wild benchmarks such as VoxCeleb~\cite{nagrani2017voxceleb} and VoxCeleb2~\cite{chung2018voxceleb2} have advanced speaker modeling from unconstrained videos, while recent studies further explore cross-modal discriminative learning and attention-based interaction beyond simple late fusion~\cite{tao2020crossmodal,praveen2024dynamic,peng2024fuse}.

Despite this progress, robust audio-visual speaker identification remains challenging in realistic polyglot scenarios. Visual cues may be degraded or unavailable due to occlusion, detection failure, or poor alignment, while acoustic cues can be affected by noise, channel distortion, or short speech duration. Such quality variations weaken fixed fusion strategies that assume both modalities are reliable. Moreover, cross-language conditions may further degrade recognition, as speaker-discriminative acoustic patterns can be entangled with phonetic and language-dependent variations, even with strong audio backbones such as ECAPA-TDNN~\cite{desplanques2020ecapa}. Therefore, existing methods may struggle with missing or quality-varying modalities in polyglot speaker identification~\cite{praveen2024dynamic,peng2024fuse,zhang2024lowquality}.

\begin{figure}[t]
    \centering
    \includegraphics[width=\linewidth]{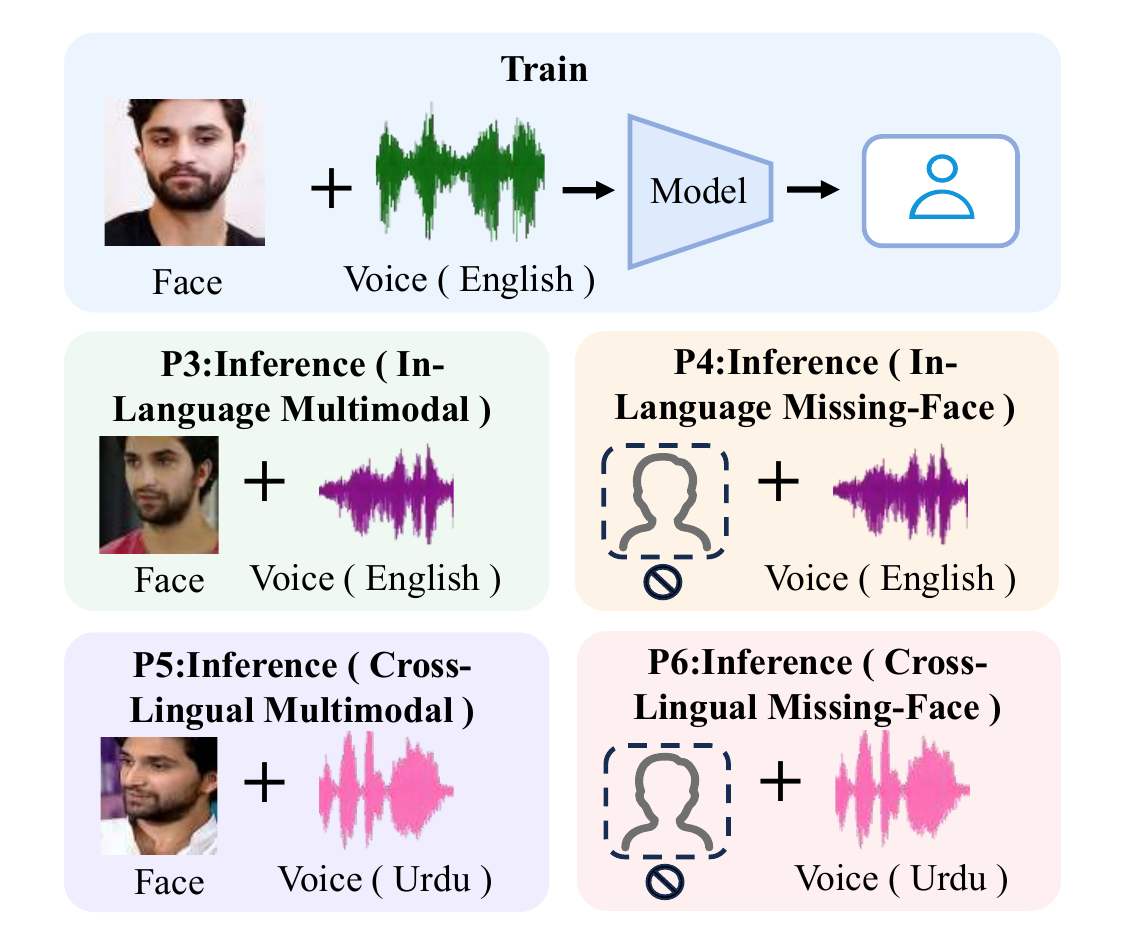}
    \caption{Task settings of the POLY-SIM 2026 Challenge, covering multimodal, missing-face, and cross-lingual speaker identification.}
    \label{fig:intro_task}
\end{figure}

These challenges are systematically formulated in the POLY-SIM 2026 Challenge~\cite{moscati2026polysim}, which targets robust speaker identification under polyglot and incomplete-modality conditions. As shown in Fig.~\ref{fig:intro_task}, the benchmark evaluates both in-language and cross-lingual speaker identification, under either complete audio-visual inputs or audio-only inputs with the face modality missing. Therefore, the task requires a unified model that can exploit complementary face-audio cues when both modalities are available, while maintaining reliable predictions when visual information is absent.

To address these challenges, we propose \textbf{MRAF}, a \textbf{M}issing-Token Prompted \textbf{R}eliability-\textbf{A}ware \textbf{F}usion framework for robust polyglot speaker identification. MRAF introduces a learnable missing token to represent unavailable face inputs, enabling a consistent token-level representation under both complete and missing-face conditions. It further estimates sample-wise modality reliability and uses reliability-aware cross-attention to adaptively fuse face and audio cues. To reduce the gap between multimodal training and audio-only inference, MRAF employs audio-only knowledge distillation. In the official POLY-SIM 2026 final evaluation, MRAF achieved 100\% accuracy on P3 and P5, obtained competitive results on P4 and P6, and ranked second overall.

Our contributions are summarized as follows:
\begin{itemize}
    \item We propose a learnable \emph{missing token} to model unavailable face inputs and maintain a unified token-level representation path under complete and missing-face conditions.
    \item We design a \emph{reliability-aware cross-attention fusion} module that adaptively modulates face-audio interaction according to sample-wise modality reliability.
    \item We introduce audio-only knowledge distillation to bridge multimodal training and missing-face inference, and validate MRAF in the POLY-SIM 2026 Challenge, where it achieves 100\% accuracy on P3/P5 and ranks second overall.
\end{itemize}

\begin{figure*}[t]
    \centering
    \includegraphics[width=\textwidth]{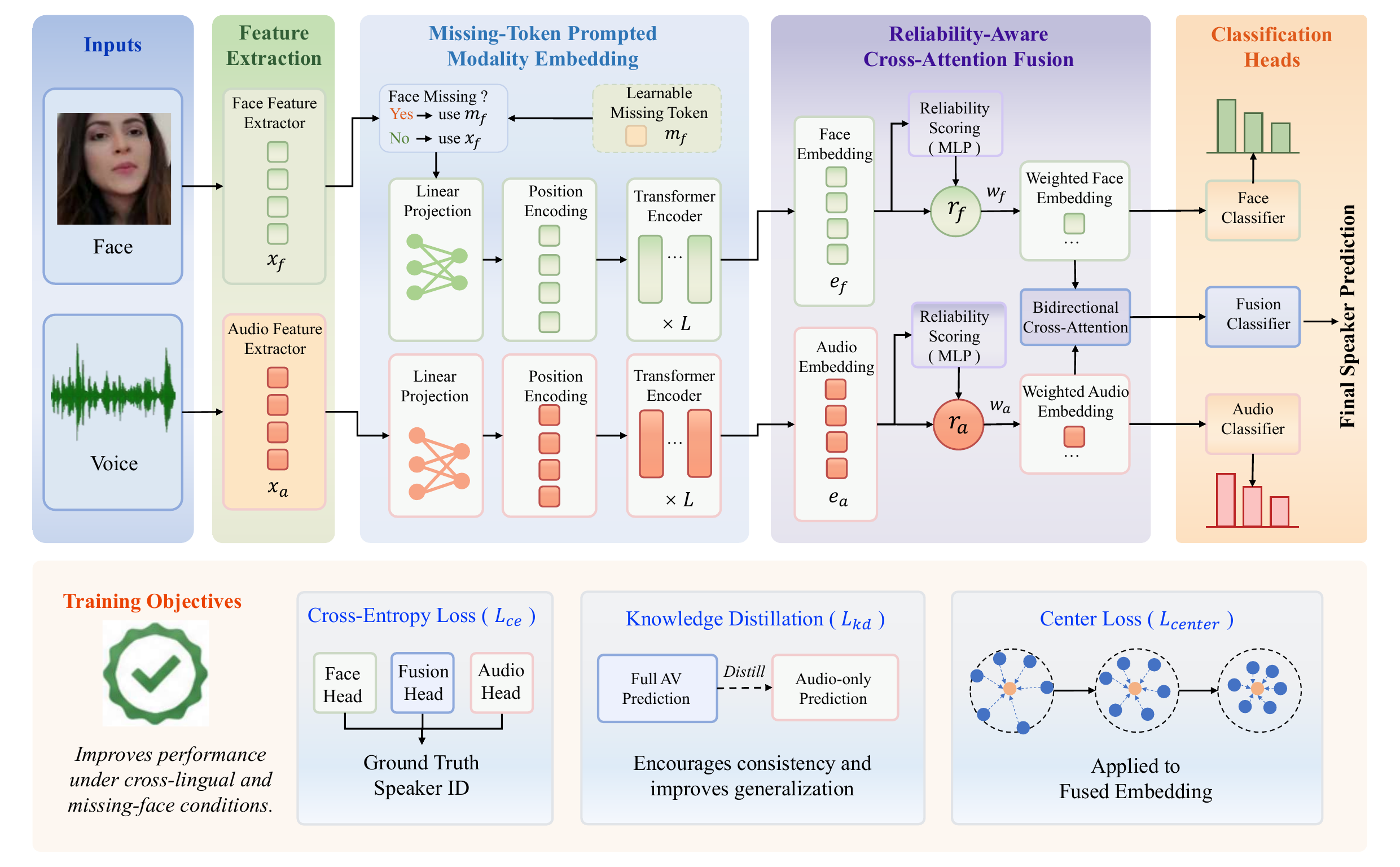}
    \caption{Overall architecture of MRAF. Pre-extracted face and audio features are projected into modality-specific token sequences, where a learnable missing token represents unavailable face inputs. Reliability scores are estimated to weight modality tokens before bidirectional cross-attention fusion. During training, face, audio, and fusion classification heads provide multi-branch supervision, while the fusion head is used for final speaker prediction during inference.}
    \label{fig:model_architecture}
\end{figure*}

\section{Related Work}


\textbf{Audio-Visual Speaker Identification.}
Audio-visual speaker identification exploits complementary facial and speech cues. VoxCeleb~\cite{nagrani2017voxceleb} and VoxCeleb2~\cite{chung2018voxceleb2} have advanced speaker recognition and audio-visual identity modeling, while MAV-Celeb and FAME 2024/2026 study multilingual face-voice association in cross-lingual environments~\cite{saeed2024synopsis,moscati2026linking}. Recent facial representation studies extend static image models to dynamic facial understanding~\cite{10663980,11207542}. Fusion strategies include joint cross-attention~\cite{praveen2023audio}, dynamic cross-attention~\cite{praveen2024dynamic}, comparative modality fusion~\cite{farhadipour2024comparative}, gated fusion~\cite{arevalo2017gated}, tensor fusion~\cite{zadeh2017tensor}, low-rank fusion~\cite{liu2018efficient}, and Multimodal Transformer~\cite{tsai2019multimodal}. Face-voice association methods, such as SBNet~\cite{saeed2023single}, Fuse after Align~\cite{peng2024fuse}, PAEFF~\cite{hannan2025paeff}, RFOP~\cite{hannan2026rfop}, XM-ALIGN~\cite{fang2026xm}, inductive-bias based class separation~\cite{moscati2026face}, deep latent cross-modal mapping~\cite{nawaz2019deep}, and shared multimodal embedding learning~\cite{simic2026shared}, focus on alignment, projection, and shared identity spaces. FuseMoE~\cite{han2024fusemoe} and Flex-MoE~\cite{yun2024flex} support arbitrary modality combinations. Unlike fixed interaction schemes, our method performs reliability-aware fusion according to modality reliability.

\textbf{Language-Robust Speaker Recognition.}
Language-robust speaker recognition seeks to maintain speaker discriminability under multilingual or language-mismatched conditions. Classical x-vector~\cite{snyder2018x} and ECAPA-TDNN~\cite{desplanques2020ecapa} perform strongly, yet multilingual speech introduces phonetic, pronunciation, and acoustic variations that weaken cross-lingual generalization. Benchmarks such as TidyVoice~\cite{farhadipour2026tidyvoice}, the TidyVoice 2026 Challenge~\cite{farhadipour2026tidyvoicechallenge}, and SVeritas~\cite{baali2025sveritas} provide systematic multilingual verification protocols. Existing methods mitigate language interference via language-invariant learning~\cite{li2026language}, Dual-LoRA adversarial disentanglement~\cite{shangguan2026dual}, Whisper-based representations~\cite{zhao2024whisper,emon2025whisper}, and phonetic-guided modeling~\cite{ma2025expo}. Multimodal speaker recognition has also been explored for realistic scenarios~\cite{shah2023speaker}, but most studies remain audio-centric and underuse visual identity cues. Our method transfers audio-visual knowledge to audio-only inference through multimodal distillation.

\textbf{Missing Modality Learning for Robust Multimodal Recognition.}
Missing modality learning addresses multimodal recognition when modalities are unavailable due to occlusion, sensor failure, privacy constraints, or corruption. Existing studies cover reconstruction, shared-specific representation learning, knowledge distillation, prompt learning, contrastive learning, and flexible fusion~\cite{wu2024deep,liaqat2025multimodal}. ShaSpec~\cite{wang2023multi} decomposes representations into shared and modality-specific components, while cross-modal alignment and reconstruction~\cite{sun2025enhancing} reduce complete-incomplete gaps. M3AE~\cite{liu2023m3ae} improves robustness via modality- or patch-level masking. Distillation methods, including LCKD~\cite{wang2023learnable}, one-stage modality distillation~\cite{wei2023one}, and Knowledge Bridger~\cite{ke2025knowledge}, transfer complete-modality knowledge to incomplete inputs. Prompt-based methods, such as multimodal prompt learning~\cite{guo2024multimodal}, deep correlated prompting~\cite{hu2024deep}, and PROMISE~\cite{chen2026promise}, adapt representations to modality availability or representation shift. Single-branch embedding has also been explored for cold-start and missing-modality scenarios~\cite{ganhor2024multimodal}. Unlike general missing-modality settings, missing-face modality removes strong identity cues, and zero filling may cause distribution shift. We use a learnable missing token as a trainable placeholder for unavailable face inputs.

\section{Method}

\subsection{Task Formulation}

The task is formulated as polyglot multimodal speaker identification. Each sample is represented by pre-extracted face and audio features, denoted as $x_f \in \mathbb{R}^{d_f}$ and $x_a \in \mathbb{R}^{d_a}$, respectively, and is associated with a language domain $\ell \in \mathcal{L}$. The objective is to predict the speaker label $y \in \{1,2,\dots,C\}$ from a closed set of $C$ speakers. In practical polyglot scenarios, the model is expected to generalize across different language domains while handling varying modality availability during inference. In particular, the visual modality may be unavailable due to occlusion. Therefore, the model needs to handle both complete face-audio inputs and audio-only inputs.

Specifically, under the complete-modality condition, both $x_f$ and $x_a$ are available. Under the audio-only condition, the face feature is replaced by a zero vector $x_f=\mathbf{0}$. The model learns a unified prediction function:
\begin{equation}
\hat{y} = \arg\max_c p(y=c \mid x_f, x_a),
\end{equation}
where the same model is applied to both complete and missing-face inputs across different language domains.

\subsection{Overall Architecture}

The proposed framework, illustrated in Figure~\ref{fig:model_architecture}, consists of three main components: the \textbf{Missing-Token Prompted Modality Embedding} module, the \textbf{Reliability-Aware Cross-Attention Fusion} module, and a multi-branch classification module. The pre-extracted face feature $x_f$ and audio feature $x_a$ are projected into modality-specific token sequences, with a learnable missing token $m_f$ representing the missing visual state when the face modality is absent. Each token sequence is encoded to produce the face embedding $e_f$ and audio embedding $e_a$, each with 512 dimensions. These embeddings are then input to the reliability-aware fusion module, where lightweight reliability scorers estimate modality confidence scores $r_f$ and $r_a$, which are normalized into weights $w_f$ and $w_a$ to modulate token-level modality contributions. The weighted tokens are further fused via bidirectional cross-attention to generate the fused embedding $e_c$. Finally, three classification heads output $z_f$, $z_a$, and $z_c$ for the face, audio, and fusion embeddings, respectively, with the fusion classifier serving as the primary prediction head during inference to produce the final speaker prediction $\hat{y}$.

\subsection{Missing-Token Prompted Modality Embedding}

The missing-token prompted modality embedding module maps pre-extracted face and audio features into a unified token-level space, producing compact modality-specific speaker representations and a learnable representation for unavailable face inputs. For each modality $m \in \{f,a\}$, the input feature $x_m \in \mathbb{R}^{d_m}$ is first projected into a sequence of $K$ latent tokens with dimension $d$:
\begin{equation}
T_m = \text{reshape}(W_m x_m + b_m),
\end{equation}
and a positional embedding $P_m \in \mathbb{R}^{K \times d}$ is added:
\begin{equation}
\tilde{T}_m = T_m + P_m.
\end{equation}

The token sequence $\tilde{T}_m \in \mathbb{R}^{K \times d}$ is then processed by $L$ Transformer encoder~\cite{vaswani2017attention} layers. Let $H_m^{(0)}=\tilde{T}_m$. In the $l$-th layer, the input sequence $H_m^{(l-1)}$ is projected into queries, keys, and values, denoted as $Q_m^{(l)}$, $K_m^{(l)}$, and $V_m^{(l)}$, respectively. For each attention head $h$, the attention output is computed as:
\begin{equation}
\text{head}_h =
\text{softmax}
\left(
\frac{Q_{m,h}^{(l)} {K_{m,h}^{(l)}}^\top}{\sqrt{d_h}}
\right)
V_{m,h}^{(l)},
\end{equation}

where $Q_{m,h}^{(l)}$, $K_{m,h}^{(l)}$, and $V_{m,h}^{(l)}$ are the split query, key, and value projections for head $h$, $d_h=d/H$ is the per-head dimension, and $H$ is the number of heads. Here, $\text{MHA}(\cdot)$ denotes the multi-head self-attention operation. The outputs of all heads are concatenated and linearly projected:
\begin{equation}
\text{MHA}(H_m^{(l-1)})
=
\text{Concat}(\text{head}_1,\dots,\text{head}_H)W_O.
\end{equation}

The layer output includes a residual connection and layer normalization, followed by a position-wise feed-forward network:
\begin{equation}
U_m^{(l)}
=
\text{LayerNorm}
\left(
H_m^{(l-1)} + \text{MHA}(H_m^{(l-1)})
\right),
\end{equation}
\begin{equation}
H_m^{(l)}
=
\text{LayerNorm}
\left(
U_m^{(l)} + \text{FFN}(U_m^{(l)})
\right),
\end{equation}
where $\text{FFN}(x)=\text{ReLU}(xW_1+b_1)W_2+b_2$. After $L$ layers, the encoded token sequence $H_m^{(L)} \in \mathbb{R}^{K \times d}$ is obtained. Finally, mean pooling across the $K$ tokens, followed by layer normalization and $\ell_2$ normalization, produces the modality embedding:
\begin{equation}
e_m =
\text{Norm}
\left(
\text{LayerNorm}
\left(
\frac{1}{K}\sum_{i=1}^{K} H_m^{(L,i)}
\right)
\right),
\end{equation}
where $e_m \in \mathbb{R}^{512}$ for each modality.

\paragraph{Missing face handling}
When the face modality is unavailable, the zero-valued face input is replaced by a learnable missing token $m_f$:
\begin{equation}
x_f =
\begin{cases}
m_f, & \text{if } x_f = \mathbf{0},\\
x_f, & \text{otherwise}.
\end{cases}
\end{equation}

This design provides a trainable and modality-aware representation for absent face information, rather than treating the missing modality as a fixed zero vector. By representing missing faces in the same token space as valid face features, the model can reduce the distribution gap caused by missing inputs and provide more informative token representations for subsequent reliability estimation and cross-attention fusion.

\subsection{Reliability-Aware Cross-Attention Fusion}

Face and audio modalities provide complementary speaker cues, but their reliability may vary across samples due to missing or degraded inputs. To address this issue, the reliability-aware cross-attention fusion module estimates the confidence of each modality and uses it to adaptively control its contribution.

Given the face embedding $e_f$ and audio embedding $e_a$, two lightweight reliability scorers are used to estimate scalar reliability scores:
\begin{equation}
r_f = \sigma(W_f^r e_f + b_f^r), \quad
r_a = \sigma(W_a^r e_a + b_a^r),
\end{equation}
where $\sigma(\cdot)$ denotes the sigmoid function. The reliability scores are then normalized into modality weights:
\begin{equation}
w_f = \frac{r_f}{r_f + r_a + \epsilon}, \quad
w_a = \frac{r_a}{r_f + r_a + \epsilon},
\end{equation}
where $\epsilon$ is a small constant for numerical stability. When the face modality is missing, its reliability score is set to zero, encouraging the fusion module to rely more on the audio modality. The normalized weights are applied to the corresponding token representations:
\begin{equation}
\hat{H}_f = w_f H_f, \quad \hat{H}_a = w_a H_a,
\end{equation}
where $H_f$ and $H_a$ denote the encoded face and audio token sequences, respectively. The reliability-weighted tokens are then fused through bidirectional cross-attention:
\begin{equation}
C_f = \text{MHA}(\hat{H}_f, \hat{H}_a, \hat{H}_a), \quad
C_a = \text{MHA}(\hat{H}_a, \hat{H}_f, \hat{H}_f),
\end{equation}
This bidirectional design allows each modality to attend to the other modality while being modulated by its estimated reliability. The attention outputs are further processed with residual connections and layer normalization, producing $\bar{C}_f$ and $\bar{C}_a$.

Finally, the fused representation is obtained by aggregating the two cross-attended token sequences:
\begin{equation}
e_c = \text{LayerNorm}\left(\frac{1}{2}\big(\text{MeanPool}(\bar{C}_f) + \text{MeanPool}(\bar{C}_a)\big)\right).
\end{equation}

The fused embedding $e_c$ serves as the primary representation for the fusion classifier, integrating face-audio cues under reliability-aware modulation. By emphasizing reliable information while suppressing unreliable cues, it supports robust speaker prediction under both complete-modality and missing-face conditions.

\subsection{Training Objectives}

The model is trained with complementary objectives to improve speaker discrimination under both complete-modality and missing-face conditions. Three classification heads are attached to the face embedding $e_f$, audio embedding $e_a$, and fused embedding $e_c$, producing logits $z_f=g_f(e_f)$, $z_a=g_a(e_a)$, and $z_c=g_c(e_c)$, respectively. The face and audio branches provide auxiliary supervision, while the fusion branch serves as the primary prediction branch. The classification loss is defined as:
\begin{equation}
\mathcal{L}_{ce}
=
\lambda_f \operatorname{CE}(z_f,y)
+
\lambda_a \operatorname{CE}(z_a,y)
+
\lambda_c \operatorname{CE}(z_c,y),
\end{equation}
where $\lambda_f$, $\lambda_a$, and $\lambda_c$ control the contributions of different branches.

To improve missing-face robustness, audio-only knowledge distillation~\cite{hinton2015distilling} is introduced by using the complete face-audio prediction as the teacher and the audio-only prediction as the student. Let $z_c^{AV}$ and $z_c^{A}$ denote the fusion logits under complete-modality and audio-only inputs, respectively. The distillation loss is:
\begin{equation}
\mathcal{L}_{kd} = T^2 \, \operatorname{KL}\Big(\operatorname{softmax}\big(z_c^{AV}/T\big) \, \| \, \operatorname{softmax}\big(z_c^{A}/T\big)\Big),
\end{equation}
where $T$ is the temperature parameter. This objective transfers discriminative knowledge from complete audio-visual inference to audio-only inference.

Center loss~\cite{wen2016discriminative} is further applied to the fused embedding to enhance intra-class compactness, defined as $\mathcal{L}_{center}=\frac{1}{2}\sum_{i=1}^{B}\|e_c^{(i)}-\mu_{y_i}\|_2^2$, where $\mu_{y_i}$ is the class center of speaker $y_i$. The overall objective is:
\begin{equation}
\mathcal{L}
=
\mathcal{L}_{ce}
+
\lambda_{kd}\mathcal{L}_{kd}
+
\lambda_{center}\mathcal{L}_{center},
\end{equation}
where $\lambda_{kd}$ and $\lambda_{center}$ balance the distillation and center losses.

\section{Experiments}

\subsection{Dataset and Evaluation Protocol}

Experiments are conducted on the MAV-Celeb dataset~\cite{nawaz2021cross} adopted by the POLY-SIM 2026 challenge~\cite{moscati2026polysim}. MAV-Celeb contains paired face images and speech audio from English and Urdu speakers collected from unconstrained YouTube videos, with each sample annotated by a speaker ID. The dataset includes diverse real-world variations in both visual and audio modalities.\footnote{Following the POLY-SIM 2026 protocol, only the English training split is used for training, while the Urdu split is reserved for cross-lingual evaluation.} Its statistics are summarized in Table~\ref{tab:dataset_split}.

\begin{table}[t]
\centering
\caption{Statistics of the MAV-Celeb dataset.}
\label{tab:dataset_split}
\begin{tabularx}{\columnwidth}{c c c}
\toprule
\textbf{Language} & \makecell{\textbf{Videos} \\ \textbf{(Train / Val / Test)}} & \makecell{\textbf{Samples} \\ \textbf{(Train / Val / Test)}} \\
\midrule
English & 262 / 70 / 70 & 4039 / 1290 / 1521 \\
Urdu    & 415 / 70 / 70 & 9304 / 1779 / 1623 \\
\bottomrule
\end{tabularx}
\end{table}

Following the challenge protocol~\cite{moscati2026polysim}, evaluation covers four settings. P3 and P5 use complete audio-visual inputs, while P4 and P6 evaluate audio-only recognition with the face modality missing. P3/P4 are in-language settings, whereas P5/P6 assess cross-lingual generalization. Top-1 accuracy is reported for each setting, and the final score is the average over all four tasks.

\subsection{Experimental Setup}

All experiments use the pre-extracted features provided by the POLY-SIM 2026 challenge~\cite{moscati2026polysim}. Face embeddings are extracted by FaceNet~\cite{schroff2015facenet}, and audio embeddings by ECAPA-TDNN~\cite{desplanques2020ecapa}. Training and ablations are performed on a single NVIDIA RTX 4090 GPU.

The model is trained with Adam~\cite{kingma2014adam}, using a learning rate of $1\times10^{-4}$, batch size 64, embedding dimension 512, and dropout 0.1. To simulate missing-face conditions, complete audio-visual and audio-only samples are used with probabilities $p_{av}=0.8$ and $p_a=0.2$, respectively. The branch loss weights are $\lambda_f=\lambda_a=\lambda_c=1.0$. The distillation temperature and weight are set to $T=2.0$ and $\lambda_{kd}=0.2$, and the center loss weight is $\lambda_{center}=0.1$.

\subsection{Comparison with Competitors}

We evaluate MRAF on the official POLY-SIM 2026 test set. As shown in Table~\ref{tab:competition_results}, MRAF ranks second overall with an average accuracy of 0.99568. Under complete-modality settings, MRAF achieves 1.00000 accuracy on both P3 and P5, obtaining the best result on P3 and tied-best result on P5. This strong performance is mainly attributed to the complementary identity cues from face and audio modalities, where visual embeddings provide stable speaker-specific information and audio embeddings capture vocal characteristics. For the more challenging missing-face settings, MRAF achieves 0.98948 on P4 and 0.99320 on P6, showing competitive robustness when only audio information is available. The remaining performance gap mainly comes from the absence of visual identity cues and the increased sensitivity of audio-only recognition to pronunciation variation, acoustic conditions, and cross-lingual shifts. Compared with the official baseline, MRAF improves the average accuracy from 0.73373 to 0.99568, with large gains of 0.46417 on P4 and 0.55451 on P6. These improvements demonstrate the effectiveness of the missing-token prompted reliability-aware fusion framework for incomplete-modality speaker identification.

\begin{table}[t]
\centering
\caption{Quantitative comparison with top-ranked submissions. Avg. denotes the mean Top-1 accuracy over P3--P6.}
\label{tab:competition_results}
\renewcommand{\arraystretch}{1.0} 
\setlength{\tabcolsep}{1pt}       
\small 
\begin{tabular}{clccccc}
\toprule
\textbf{Rank} & \textbf{Team} & \textbf{Avg.} & \textbf{P3} & \textbf{P4} & \textbf{P5} & \textbf{P6} \\
\midrule
1 & Ayoub ELKHOUZARI & 0.99886 & 0.99803 & 0.99803 & 1.00000 & 0.99938 \\
2 & \textbf{MRAF (Ours)} & \textbf{0.99568} & \textbf{1.00000} & \textbf{0.98948} & \textbf{1.00000} & \textbf{0.99322} \\
3 & tartarz & 0.99066 & 0.99934 & 0.97502 & 1.00000 & 0.98829 \\
4 & areffarhadi & 0.98776 & 0.99934 & 0.96647 & 0.99938 & 0.98583 \\
5 & Yassin TERRAF & 0.98324 & 0.99934 & 0.96318 & 0.99322 & 0.97720 \\
6 & rin\_thira & 0.98258 & 0.99737 & 0.96252 & 0.99938 & 0.97104 \\
7 & whalemyj & 0.97600 & 0.99934 & 0.94346 & 0.99815 & 0.96303 \\
8 & mmosc (Baseline) & 0.73373 & 0.98817 & 0.52531 & 0.98275 & 0.43869 \\
\bottomrule
\end{tabular}
\end{table}

\begin{figure*}[t]
    \centering
    \includegraphics[width=\textwidth]{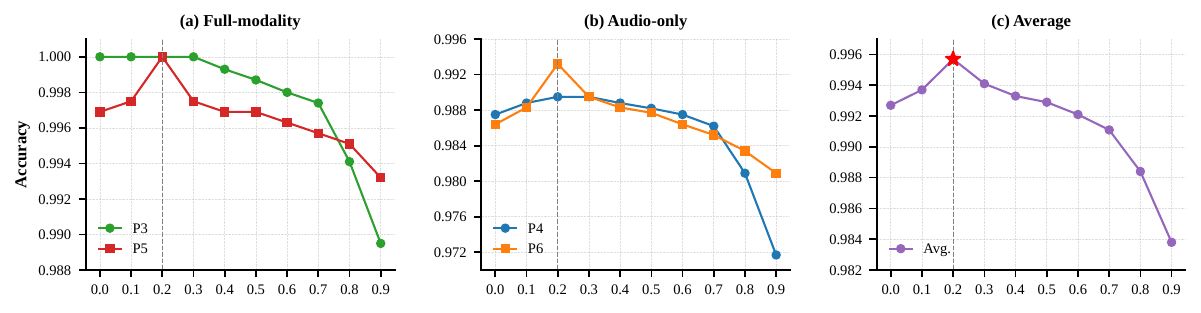}
    \caption{Effect of the sampling ratio between full-modality and audio-only training samples. 
    The x-axis denotes the audio-only sampling probability $p_a$.
    Results are shown for (a) full-modality settings P3 and P5, 
    (b) audio-only settings P4 and P6, and 
    (c) average accuracy across all four settings. 
    The best average performance is achieved at $p_a=0.2$ and $p_{av}=0.8$.}
    \label{fig:ablation_ratio}
\end{figure*}

\subsection{Ablation Study}

\subsubsection{Ablation on Fusion Strategy}

We first evaluate the effectiveness of the proposed reliability-aware cross-attention fusion module by comparing it with linear fusion, gated fusion, and LSTM-based fusion. As shown in Table~\ref{tab:ablation_fusion}, cross-attention achieves the best average accuracy of 0.9957, outperforming linear, gated, and LSTM-based fusion by 0.1264, 0.1324, and 0.0022, respectively. Linear and gated fusion mainly combine modality features through fixed or sample-level weights, which limits their ability to model fine-grained face-audio interactions. In contrast, cross-attention enables token-level bidirectional interaction between modalities and adaptively emphasizes more reliable cues, leading to stronger robustness, especially in the missing-face settings P4 and P6. Although LSTM-based fusion performs competitively, it models the fused sequence in a relatively implicit manner and is slightly less effective than direct cross-modal attention.

\begin{table}[t]
\centering
\caption{Ablation study on different fusion strategies.}
\label{tab:ablation_fusion}
\setlength{\tabcolsep}{4pt}
\renewcommand{\arraystretch}{1.05}
\begin{tabular}{cccccc}
\toprule
\textbf{Fusion Method} & \textbf{P3} & \textbf{P4} & \textbf{P5} & \textbf{P6} & \textbf{Avg.} \\
\midrule
Linear & 0.8974 & 0.9020 & 0.8349 & 0.8429 & 0.8693 \\
Gated & 0.8679 & 0.8527 & 0.8681 & 0.8644 & 0.8633 \\
LSTM & 0.9987 & 0.9882 & 0.9975 & 0.9895 & 0.9935 \\
\textbf{Cross-attention} & \textbf{1.0000} & \textbf{0.9895} & \textbf{1.0000} & \textbf{0.9932} & \textbf{0.9957} \\
\bottomrule
\end{tabular}
\end{table}

\subsubsection{Ablation on Missing-Modality Handling}

We then compare different strategies for handling missing face inputs, including zero filling, audio-based completion, audio-guided memory bank selection, and the proposed learnable missing token. As reported in Table~\ref{tab:ablation_missing}, zero filling provides a simple baseline but introduces an artificial feature pattern that is not aligned with the distribution of real face embeddings. Audio completion and memory bank selection provide more informative substitutes, but they rely on heuristic reconstruction or retrieval and may introduce inaccurate visual cues. The learnable missing token achieves the best average accuracy of 0.9957 and improves P6 from 0.9901 to 0.9932 over zero filling. This suggests that a trainable missing representation can better encode the absence of face modality and provide a stable placeholder for cross-modal fusion.

\begin{table}[t]
\centering
\caption{Ablation study on different missing-modality handling strategies.}
\label{tab:ablation_missing}
\setlength{\tabcolsep}{3pt}
\renewcommand{\arraystretch}{1.05}
\begin{tabular}{lccccc}
\toprule
\textbf{Method} & \textbf{P3} & \textbf{P4} & \textbf{P5} & \textbf{P6} & \textbf{Avg.} \\
\midrule
Zero filling & 0.9993 & 0.9882 & 0.9982 & 0.9901 & 0.9940 \\
Audio completion & 0.9993 & \textbf{0.9895} & 0.9975 & 0.9895 & 0.9940 \\
Memory bank & 0.9993 & \textbf{0.9895} & 0.9982 & 0.9914 & 0.9946 \\
\textbf{Learnable token} & \textbf{1.0000} & \textbf{0.9895} & \textbf{1.0000} & \textbf{0.9932} & \textbf{0.9957} \\
\bottomrule
\end{tabular}
\end{table}

\subsubsection{Ablation on Full-Modality and Audio-Only Sampling Ratio}

We further study the effect of the training sampling ratio between complete audio-visual samples and audio-only samples, where $p_a$ denotes the probability of audio-only samples and $p_{av}=1-p_a$ denotes that of full-modality samples. As shown in Figure~\ref{fig:ablation_ratio}, using only full-modality samples yields strong performance on complete-modality evaluation but lower robustness under missing-face conditions. Introducing audio-only samples improves P4 and P6 by exposing the model to missing-modality patterns during training. The best average performance is obtained at $p_a=0.2$ and $p_{av}=0.8$, indicating a good balance between exploiting visual identity cues and improving audio-only inference. When $p_a$ becomes too large, performance decreases because excessive audio-only training weakens the model's ability to benefit from visual information.

\subsubsection{Limitations and Failure Cases}

Although the proposed method achieves high accuracy, it may still fail under noisy utterances, low-quality face images, or face-audio mismatch, where reliable speaker cues are weakened. Stronger cross-lingual shifts in pronunciation, phonetic patterns, or recording conditions may also limit generalization beyond the current evaluation setting.

\section{Conclusion}

In this paper, we presented MRAF for robust polyglot speaker identification with missing-face modality. Rather than using zero-filled visual features, MRAF introduces a learnable missing token to represent unavailable face information, enabling complete and missing-face inputs to share a unified token-level modeling process. Reliability-aware cross-attention adjusts face-audio interaction according to sample-wise modality confidence, improving robustness when one modality is unreliable. Audio-only knowledge distillation narrows the gap between multimodal training and missing-face inference, while center loss encourages compact speaker representations. Experiments on POLY-SIM 2026 show that MRAF performs strongly across complete-modality, missing-face, and cross-lingual settings, ranking second overall. Ablation studies confirm the contributions of cross-attention fusion, the missing token, and the balance between full-modality and audio-only training samples. Future work will explore more general missing-modality settings and stronger language-invariant speaker representations.


\begin{acks}
This work was supported in part by the National Key Research and Development Program under Grant No. 2023YFC2506803, and in part by the Fundamental Research Funds for the Central Universities of China under Grant No. PA2025IISL0110. The computations were performed on the High-Performance Computing (HPC) Platform of Hefei University of Technology.
\end{acks}

\bibliographystyle{ACM-Reference-Format}
\bibliography{sample-base}

\end{document}